# Assessing the detection of the Yarkovsky effect using Gaia DR3 and FPR catalogues

Karolina Dziadura[1,3], Przemysław Bartczak[1,2], and Dagmara Oszkiewicz[1]

[1] Astronomical Observatory Institute, Faculty of Physics, A. Mickiewicz University, Słoneczna 36, 60-286 Poznań, Poland
[2] Instituto Universitario de Física Aplicada a las Ciencias y las Tecnologías (IUFACyT). Universidad de Alicante, Ctra. San Vicente del Raspeig s/n. 03690 San Vicente del Raspeig, Alicante, Spain
[3] Harvard-Smithsonian Center for Astrophysics, 60 Garden St., MS 15, Cambridge, MA, 02138, USA

November 14, 2024

**ABSTRACT**

*Aims.* The primary objective of this study was to utilize the newest *Gaia* FPR catalogue containing ultra-precise asteroid astrometry spanning over 66 months to detect the Yarkovsky effect, a non-gravitational acceleration that affects the orbits of small asteroids. Moreover, this study examines close approaches of near-Earth asteroids by comparing orbits calculated using the *Gaia* data.
*Methods.* We used the conventional least-squares orbit computation method available in the OrbFit software (version developed at the MPC). We used the latest *Gaia* Focused Product Release, complemented by data from the Minor Planet Center and radar astrometry from the Jet Propulsion Laboratory. We fitted orbital parameters for 446 Near-Earth Asteroids, including the additional non-gravitational transverse acceleration to model the Yarkovsky effect. Furthermore, we compared the results obtained using different datasets: *Gaia* Focused Product Release and the previous *Gaia* Data Release 3.
*Results.* We detected a robust Yarkovsky effect in 43 NEAs. As expected, we found an improvement in the orbital elements uncertainty and in the signal-to-noise ratio of the Yarkovsky effect detections when using the current *Gaia* FPR with double the observing arc compared to the DR3 catalog. We also found nine additional reliable detections of the Yarkovsky effect when using the new *Gaia* FPR catalog.
*Conclusions.* Including the Yarkovsky effect in the force model can be important to reliably estimate close approach distances of near-Earth asteroids. Several of the detected Yarkovsky drifts already have a signal-to-noise ratio greater than 10, which is high enough for their Yarkovsky effect to be included in their reliable long-term orbital evolution, close approach, and Earth impact analysis. The final *Gaia* catalog may provide a much higher number of high signal-to-noise ratio detections of the Yarkovsky effect. This would allow a broader inclusion of the effect in routine close approach and impact analysis of these objects.

## 1. Introduction

The Gaia mission is a revolutionary mission of the European Space Agency that delivers exceptionally precise astrometric measurements of celestial bodies throughout the entire sky (Prusti et al., 2016). Although the primary objective is to generate a precise 3D map of stars in the Milky Way, the mission also measures accurate positions of asteroids, including Near-Earth Objects (NEOs). The latest data release, known as the Focus Product Data Release (FPR), contains astrometric data for 446 NEOs, with an average astrometric uncertainty of ∼ 50 mas (David et al., 2023; Tanga et al., 2022; Spoto et al., 2018).

These precise observations can increase the capability of determining the Yarkovsky effect, a phenomenon associated with the anisotropic thermal emission of solar radiation absorbed by asteroid surfaces, leading to secular changes in the orbital semi-major axis (Farinella et al., 1998; Bottke et al., 2006; Vokrouhlický et al., 2015). The effect plays a critical role in the delivery of NEOs from the main asteroid belt, long-term orbit propagation, calculation of Earth impact probabilities, impact hazard assessment, and development of asteroid deflection strategies (Bottke et al., 2006; Farnocchia et al., 2015).

The Yarkovsky effect has seasonal and diurnal components (Vokrouhlický et al., 2015). The seasonal component relates to the seasonal heating and cooling of an object's hemispheres in its orbital motion. The diurnal effect relates to the heating and emission of thermal photons during the object's rotation. The seasonal effect was first detected for the LAGEOS satellite (Rubincam, 1988) and the diurnal effect for the asteroid (6489) Golevka (Chesley et al., 2003). The overall Yarkovsky effect depends on the surface conductivity, density, shape, and heliocentric distance of the object (Bottke et al., 2006; Vokrouhlický et al., 2015). For a 1 km-sized NEO, the effect is on the order of $10^{-4}$ au/Myr (Greenberg et al., 2017)

The significance of the Yarkovsky effect was recently highlighted by the case of the asteroid (99942) Apophis, considered one of the most hazardous objects for nearly two decades prior to its 2021 observations. Precise radar measurements in 2021 allowed an accurate determination of the semimajor axis drift due to the Yarkovsky effect of d$a$/d$t$ = -199.0 ± 1.5 m / year [1] (Pérez-Hernández & Benet, 2022). This resulted in the reduction of its Earth impact probability during its 2029, 2036, and 2068 close approaches, which ultimately led to Apophis's removal from impact risk tables[2]. On 13 April 2029, during its next approach, Apophis will pass Earth at a distance smaller than the radius of the geostationary orbit. Presently, there are more than 1500 asteroids with non-zero impact probabilities listed for the nearest 100 years. The Yarkovsky effect has recently been included in routine Earth impact probability computations at the impact-monitoring system Sentry-II at JPL (Roa et al., 2021). Previously, the Yarkovsky effect was factored into impact monitoring for only four asteroids: (99942) Apophis, (410777) 2009 FD, (101955) Bennu, and (29075) 1950 DA, due to the sufficiently well-determined effect (NEODys[3]). Thanks to the NASA OSIRIS-REx mission, the Yarkovsky drift of (101955) Bennu has been determined at $da/dt = -284.6 \pm 0.2$ m / year, with an impressive signal-to-noise ratio of $\sim 1400$ (Farnocchia et al., 2021). The Yarkovsky effect for asteroids (410777) 2009 FD and (29075) 1950 DA has been established through radar and optical observations (Del Vigna et al., 2019; Farnocchia & Chesley, 2014). The estimation of the Yarkovsky drift for the 2009 FD asteroid played a crucial role in excluding its potential Earth impact in 2185 (Del Vigna et al., 2019), and for the 1950 DA, it allowed the estimate of the impact probability in the year 2880, while before that only an upper limit for its impact probability was derived (Giorgini et al., 2002). The Yarkovsky effect has already been detected for numerous other NEOs (e.g. Farnocchia et al., 2013a; Chesley et al., 2015; Nugent et al., 2012; Liu et al., 2023; Del Vigna et al., 2018; ?), but with lower statistical significance. The precise astrometry of the ESA Gaia mission is proving to be a game changer for many of these asteroids (Delbo et al., 2008).

---

[1] https://www.jpl.nasa.gov/news/nasa-analysis-earth-is-safe-from-asteroid-apophis-for-100-plus-years
[2] ESA NEOCC risk list: https://neo.ssa.esa.int/risk-list, NASA Sentry risk table: https://cneos.jpl.nasa.gov/sentry/,

In this work, we present the determination of the Yarkovsky effect for 52 NEOs contained in the *Gaia* Focused Product Release. Additionally, we conduct an analysis of close approaches for selected NEOs. Section 2 provides a description of the key features of the *Gaia* Focused Product Release astrometry and other data used in this investigation. The results are presented in Section 4, while Sections 3 and 5 cover the methods and summary, respectively.

## 2. Data

We used optical astrometry from the *Gaia* Focused Product Release (FPR) and the Minor Planet Center (MPC) as well as radar data from the Jet Propulsion Laboratory (JPL).

The European Space Agency *Gaia* mission was launched in 2013 (Prusti et al., 2016). Through systematic scanning of the sky, the mission continues to provide milliarcsecond astrometry for a variety of celestial objects, including near-Earth asteroids. The newest *Gaia* Focused Product Release (FPR) published on October 10, 2023, offers a substantial enhancement over previous releases, particularly *Gaia* Data Release 3 (DR3) (David et al., 2023; Vallenari et al., 2023; Tanga et al., 2022). The FPR includes precise astrometric data for more than 150,000 asteroids over observational arcs that span up to 66 months. The observing arcs are about twice as long as those recorded by DR3. These extended orbital arcs provide a significant improvement in the determination of orbital parameters, reducing the uncertainties of orbital elements (David et al., 2023). *Gaia* asteroid observations have different accuracy in the Along-Scan (AL) and Across-Scan (AC) directions, with the AL direction exhibiting superior precision at the submilliarcsecond level, in contrast to the AC direction, where accuracy is considerably lower, up to approximately one arcsecond. This disparity is due to the fact that AL data are derived from *Gaia*'s astrometric field, whereas AC information is solely sourced from the sky mapper field of the CCD. This leads to a significant correlation between right ascension (RA) and declination (DEC) in *Gaia* astrometry, highlighting the need to use the full observation covariance matrix for orbit determination. This approach is essential to fully exploit the precision of *Gaia* astrometric measurements (David et al., 2023; Tanga et al., 2022; Spoto et al., 2018). In *Gaia* FPR, a marked advance over *Gaia* DR3 is not only the extended observational arc, but also the comprehensive reprocessing of the data. The independent processing methodology of the FPR, which used distinct selection criteria and acceptance thresholds for residuals, has resulted in occasional differences in the data sets. FPR has led to a significant improvement in orbital determination uncertainty compared to DR3 (David et al., 2023).

Except for *Gaia* FPR astrometry, we used the complete data set available from the Minor Planet Center (MPC) on 7 November 2023 [4]. The MPC database provides a comprehensive catalog of positional information for minor planets, comets, and irregular outer natural satellites with an accuracy ranging from a few to several hundred milliarcseconds (mas). This dataset complements the Gaia data by providing a much longer observational arc essential for the long-term study of orbital evolution of non-gravitational transverse acceleration. In the realm of astrometric observations of small Solar System bodies reported to the Minor Planet Center (MPC), the accuracy can vary significantly, often ranging from a fraction of to over 1 arc second (Vereš et al., 2017). This variability in precision is influenced by factors such as the observation equipment used, the observing conditions, and the data processing methodologies employed. Consequently, the MPC astrometric data, which encompass a wide range of sources, exhibit a broad spectrum of accuracy levels, reflecting the diverse nature of the contributing observations. Furthermore, we incorporate all available radar astrometry data from the Jet Propulsion Laboratory [5]. Radar astrometry is particularly valuable for determining precise orbits of NEOs, as it provides highly accurate range and velocity measurements that are not subject to the same limitations as optical observations (Ostro et al., 2002). Integration of JPL's radar data with optical astrometry from Gaia and the MPC allows for more precise and reliable orbit determination. For this study, we used a comprehensive dataset comprising 625,531 observations from the MPC, 66,071 from *Gaia*, and 794 radar measurements. The summary of the observations used is in Table A.1.

Combining these diverse datasets-Gaia's optical astrometry, MPC's minor planet catalog, and JPL's radar measurements-enables a multifaceted approach to understanding the dynamical properties of asteroids. It is crucial for precise orbit determination, especially for potentially hazardous asteroids (PHAs), and for supporting future space missions.

## 3. Methods

### 3.1. Orbit computation

The orbits of the investigated Near-Earth Objects (NEOs) were determined through the well-established least-squares differential correction procedure described extensively (Milani & Gronchi, 2010; Milani et al., 2005; Farnocchia et al., 2015). In each iteration, adjustments were made to the six orbital elements,

---

NEODys: `https://newton.spacedys.com/neodys/index.php?pc=4.1`
[3] `https://newton.spacedys.com`

[4] `https://www.minorplanetcenter.net`
[5] `https://ssd.jpl.nasa.gov/sb/radar.html`

along with an additional parameter $A_2$ (which parametrizes the Yarkovsky drift), until the algorithm converged to the minimum of the linearized target function (Milani & Gronchi, 2010; Milani et al., 2005; Farnocchia et al., 2015), resulting in a nominal orbit. The transverse acceleration $\mathbf{a_t} = A_2 g(r)\hat{\mathbf{t}}$ is assumed to be directly related to the secular drift in semi-major axis caused by the Yarkovsky effect, and $g(r)$ is a function of the heliocentric distance $r$ of the asteroid. Following Farnocchia et al. (2013a), we use $g(r) = r^{-2} au^{-2}$. The unit vector $\hat{t}$ indicates the direction tangent to the orbit. The secular drift in semi-major axis $da/dt$ can be estimated from Gauss variational equations (Farnocchia et al., 2013a):

$$\frac{da}{dt} = \frac{2A_2(1-e^2)}{np^2} \quad (1)$$

where $e$ is the eccentricity, n is the mean motion, and p is the semilatus rectum.

The force model incorporated gravitational accelerations from the Sun, planets, Moon, 16 largest asteroids, and the non-gravitational Yarkovsky effect. We used a modified version of the OrbFit software that was developed at the Minor Planet Center [6]. The orbital solution was computed at an epoch near the midpoint of the observational arc. The OrbFit fitting program fully accounts for the gravitational deflection of light, refining the approximation used in Gaia FPR by accurately modeling the finite distances of observed objects. We selected geocentric positions from the Gaia data release, avoiding potential errors linked to the use of barycentric positions and discrepancies in planetary ephemerides (David et al., 2023). Additionally, the difference in time scales between the Gaia pipeline (TCB) and OrbFit (TDB) has been addressed by converting to TDB within OrbFit. Considering that our study focuses on smaller NEAs, the offset between the center-of-light and center-of-figure was not incorporated into our calculations. This decision reflects the limited impact of this factor on the precision required for our investigation of the Yarkovsky effect.

Before orbit computation, astrometry was corrected for stellar biases originating from various catalogues (Farnocchia et al., 2015). Additionally, an observation weighting scheme was incorporated, considering factors such as the year of observation, observatory code, apparent magnitude, type of observations, etc. High-precision Gaia astrometry was incorporated, encompassing consideration of the complete error covariance matrix, which accounts for both random and systematic errors (Spoto et al., in prep.). For a more detailed description of the fitting procedure, please see our earlier works (Dziadura et al., 2022) and (Dziadura et al., 2023). Close approaches were computed through the propagation of nominal orbits (Milani & Gronchi, 2010).

### 3.2. Acceptance criteria

To decide whether the empirically detected Yarkovsky effect is reliable, we compare the derived value with the theoretically expected. The expected value is computed on the basis of NEA's orbital and physical properties as (Spoto et al., 2015; Del Vigna et al., 2018):

$$\langle da/dt\rangle_{\exp} = (da/dt)_{\mathcal{A}} \times \left(\frac{\sqrt{a_{\mathcal{A}}}(1-e_{\mathcal{A}}^2)}{\sqrt{a}(1-e^2)}\right)\left(\frac{D_{\mathcal{A}}}{D}\right)\left(\frac{\rho_{\mathcal{A}}}{\rho}\right)\left(\frac{\cos\phi}{\cos\phi_{\mathcal{A}}}\right)\left(\frac{1-A}{1-A_{\mathcal{A}}}\right) \quad (2)$$

---
[6] http://adams.dm.unipi.it/orbfit/

Where $a$ is the orbit's semimajor axis, $e$ is eccentricity, $D$ is diameter, $\rho$ is density, $\phi$ is obliquity, A is Bond albedo, and the values denoted with $\mathcal{A}$ are (99942) Apophis properties. The parameter S is then calculated as $S = A_{2\,\text{empirical}}/A_{2\,\text{expected}}$, where $A_{2\,\text{empirical}}$ is the empirical value derived from orbital fitting and $A_{2\,\text{expected}}$ is the expected value (Eq. 2). We define accepted values as results that show S > 2 and a signal-to-noise ratio (SNR) > 3 and marginal with SNR > 2.5. These thresholds are depicted as vertical and horizontal lines, respectively, in Figure 1. Our methodology aligns with the approaches documented in (Del Vigna et al., 2018; Spoto et al., 2015; Dziadura et al., 2022, 2023) and is compatible with that recently adopted by the ESA NEO Coordination Center (?). Objects with SNR above the threshold and S above two are considered spurious detections. These detections should not occur except in cases caused by different mechanisms for non-gravitational accelerations (A2). High A2 values may result from physical mechanisms such as thermal effects and solar radiation pressure (Farnocchia et al., 2023). Furthermore, the biases from older data could contribute to the erroneous detection of non-gravitational accelerations (Farnocchia et al., 2013a). The S value is estimated based on well-known orbital parameters and poorly known physical parameters of asteroids. However, the criterion is very strict, ensuring we do not claim values that might not be true. We also analyzed the results of A2 using different observational arcs to validate our findings.

### 4. Results

We computed orbits for 446 NEAs incorporating the additional $A_2$ parameter. Of these, 43 can be considered credible and 7 probable. The detailed results for these 43 accepted $A_2$ values can be found in Table A.1. The orbital solutions for seven asteroids deemed probable are significantly influenced by a few older, separated, or precovery observations. The comparison of results with and without the inclusion of these earlier data is presented in Table A.2. These precovery detections provided additional data points, improving the orbital fits and reducing uncertainties. However, the question remains if we fully trust those few observations. Asteroid (4179) Toutatis with a result of $A_2$ = -6.26±0.60, a tumbling asteroid, was excluded from our analysis because of its sensitivity to the selected number of perturbers and errors in their masses (Farnocchia et al., 2013a; ?). Our acceptance and validation criteria involved comparing all results with the theoretically expected values of the $A_2$ parameter, which were estimated based on orbital and physical parameters as described in 3.2. We required that the Yarkovsky drift derived from the orbit fitting be no more than twice as large as the expected drift based on the object's physical and orbital properties. In addition, we required that the accepted objects have a signal-to-noise ratio SNR greater than 3 and marginal cases greater than 2.5. These thresholds are represented by vertical and horizontal lines, respectively, in Figure 1. We identified 50 accepted objects (green) and 14 marginal objects (blue). Further details can be found in Table A.1.

We highlight the impact of the new *Gaia* FPR catalog on orbit determination processes compared to *Gaia* DR3. Figure 2 illustrates the change in the accepted Yarkovsky drift objects, where the start of each gray line represents the value determined by DR3 and the end, marked by a colored point, indicates the value determined by FPR. Other parameters and observations remained unchanged. The shift between using the two catalogs is visible mostly along the SNR axis. Due to the extended observational arc in the FPR for most objects, we observed an increase in the SNR ratio. Three asteroids moved to marginal cases as a result of a small decrease in SNR. In particular, the asteroid (66391) Moshup shifted into the rejected category. This change was attributed to a minor decrease in $A_2$ in the FRP data $A_2$ = -2.7 $10^{-15}$ au d$^{-2}$, compared to the DR3 data $A_2$ = -3.7 $10^{-15}$ au d$^{-2}$, while uncertainty levels remained similar. However, a larger number of objects benefitted from *Gaia* FPR, 9 objects moved to the accepted category, and 23 objects significantly increased the SNR ratio of their detections.

Further comparison is shown in Figure 3, which displays $\sigma_{A_2}$ as a function of the arc in orbital periods covered exclusively by Gaia data for orbit determination, accompanied by the corresponding distribution plots. The difference between the distributions comes from the additional data in the FPR as opposed to the earlier DR3. For a few objects with extensive radar observations and a well-established orbit, this addition does not significantly alter the uncertainty. However, even these slight variances in orbital uncertainties have led to the identification of new detections. In particular, the overall distribution of the results differs between the datasets; in the case of FPR, this distribution is moved towards lower values of $\sigma_{A_2}$. A direct comparison of the accepted values between the two catalogs is presented in Figure A.1, where we plot all accepted objects (top panel) and marginal objects (bottom panel) using FPR (purple) and juxtapose them with the DR3 values (green). In the upper panel, any DR3 value that does not meet the acceptance criteria is indicated by a diamond-shaped point. We indicate an overall consistency between DR3 and FPR, accompanied by an improvement in the uncertainty of $A_2$. In particular, the application of FPR data reduced the uncertainty in orbital parameters in most cases and resulted in nine additional accepted detections.

In addition, we compared our results with the available $A_2$ values from the JPL database Figure 4, which show minor differences from values reported in other studies. These differences arise primarily due to the use of alternate observational datasets and the instability of the orbital solutions depending on the specific data sets utilized. The final result appears to be sensitive to the chosen weighting model, with other observational arcs potentially included in the calculations. Most of our solutions are in agreement within the 1-sigma confidence level. However, as shown in the figure, there are a few asteroids that deviate significantly, with a value $S > 2$, which led us to reject them. In the corresponding figure, we compare all the asteroids in our sample with the available $A_2$ values from the JPL database. Of the 72 asteroids, some were rejected from our analysis, and these are also shown in the comparison.

Furthermore, we computed the orbits using only the *Gaia* and radar data. Various studies (Spoto et al., 2018, 2019; Mouret & Mignard, 2011) previously proposed that it may be possible to detect the Yarkovsky effect using only the *Gaia* data. Our analysis resulted in the accepted value for only one object with *Gaia* DR3 and four objects with *Gaia* FPR. This success was largely due to the significant amount of radar observations available. However, relying solely on *Gaia* or radar data was not sufficient to detect the Yarcovsky effect. This highlights the importance of incorporating all available data sources for a precise orbit determination. The potential for detection using only the *Gaia* data may increase with the anticipated publication of the complete *Gaia* catalog in the future (Spoto et al., 2018, 2019; Mouret & Mignard, 2011).

The impact of the Yarkovsky effect on orbit predictions has been highlighted in previous studies, such as (Farnocchia et al., 2013b), where the analysis of Apophis' 2029 close approach shows significant differences in uncertainty estimates when Yarkovsky is considered versus a gravity-only model. To extend this analysis, we included the Yarkovsky effect in our computations of future close-approach distances of the 50 studied NEOs for 165 years forward starting from Nov. 2023 and ending on Nov. 2188, we then compared those distances to close-approach distances computed based on purely gravitational orbits. using orbits computed based on orbits determined using *Gaia* DR3 and FPR. For each asteroid, we computed the close approach distance based on a purely gravitational orbit and an orbit including the non-gravitational Yarkovsky effect. The differences between those distances for the 50 asteroids are shown in Figure 5 for orbits computed on the basis of two *Gaia* catalogs. Generally, the differences are consistent between the two catalogs. The differences between the distance based on purely gravitational orbit and that with the Yarkovsky effect reach up to 0.1 lunar distances for most asteroids. For a single asteroid ((164121) 2003 YT1), we found a difference of up to 10 lunar distances. (164121) The results appear highly sensitive to the Gaia data and the weighting scheme applied. We now consider this result to be accurate, as it is in agreement with the solution provided by JPL Database. Close approaches to Earth amplify the effect on an asteroid's orbit. Even a small difference in the orbital parameters can cause this effect to accumulate over time, which explains the significant differences in the close approach distances shown in Figure 5 for asteroid 164121. This asteroid passed relatively close to Earth (0.011 au in 2073), and as a result, gravitational interactions during such approaches play a significant role in shaping its orbital evolution. However, addi-tional observations will be necessary to confirm the stability of the result in the current observational framework.

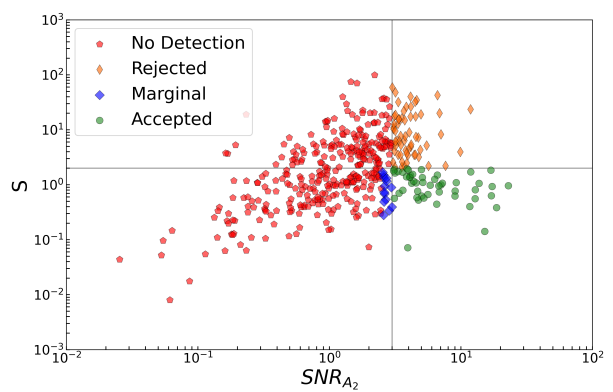

Fig. 1: Parameter S in relation to $SNR_{A_2}$ across all NEAs, as derived using *Gaia* FPR data. Green circles denote accepted values, characterized by $SNR_{A_2}$ greater than 3 and S values below 2, aligning with the scaled expected value for (99942) Apophis's $A_2$ parameter. Blue diamond indicate borderline cases where $SNR_{A_2}$ ranges between 2.5 and 3 and S remains under 2. Red stars highlight the instances where $SNR_{A_2}$ falls below 3 and orange thin diamonds represents objects with $SNR_{A_2}$ over 3 and S surpasses 2. The plot features a horizontal demarcation at S = 2 and a vertical threshold at $SNR_{A_2}$ = 3.

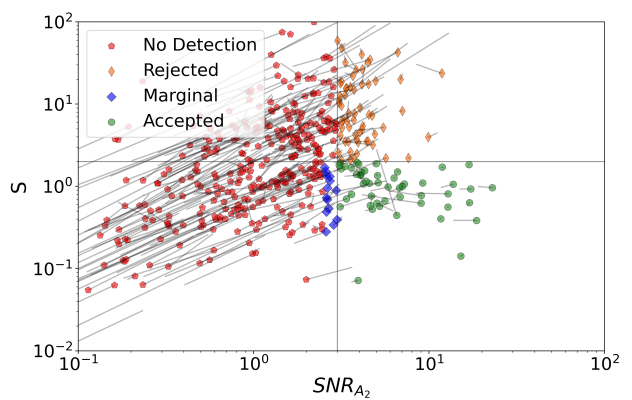

Fig. 2: Grey lines originating from the positions corresponding to *Gaia* DR3 data, leading to their new locations as determined by the application of *Gaia* FPR data. Axes and colors as in Figure 1.

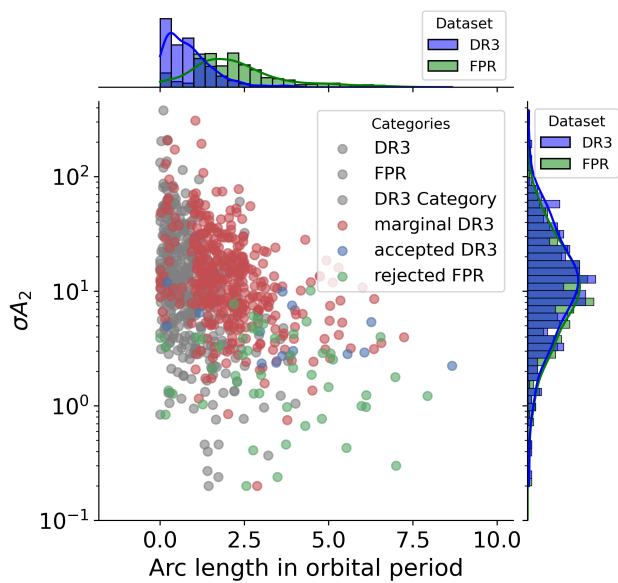

Fig. 3: Relative formal uncertainty of the Gaia $A_2$ orbital parameter as a function of the arc-length coverage expressed in orbital periods. The green, blue and red are accepted, marginal and rejected objects computed using *Gaia* FPR (density plot - green). The grey dots show the solution with the use of *Gaia* DR3 (density plot - blue).

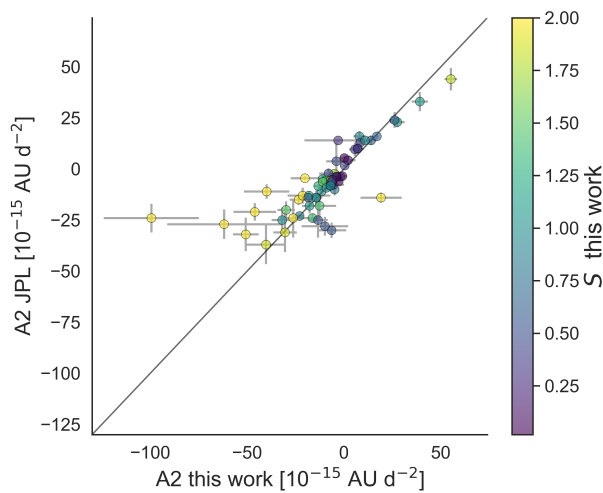

Fig. 4: Comparison of $A_2$ values between our dataset and the JPL database for 72 asteroids. Rejected objects (in this study) with $S > 2$ are yellow.

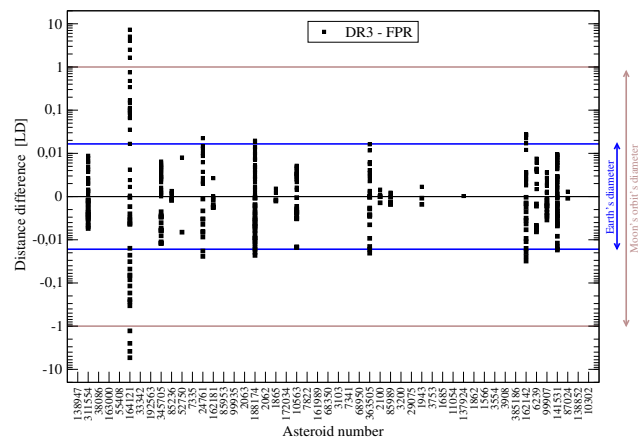

Fig. 5: The difference in distance during close approaches between distances determined from orbits without and with the Yarkovsky effect based on two different Gaia catalogues.

## 5. Summary

We searched for signal of the Yarkovsky effect among 446 NEAs, which were observed by Gaia. Of these, 50 successfully passed our validation criteria. However, 7 of these depend on old separated data. Our study confirms the crucial role of combined data sources in enhancing precision in determining the Yarkovsky effect. In particular, integration of the *Gaia* FPR catalog has demonstrated its utility in the refinement of orbital parameters. This is also evidenced by the identification of additional detections of the Yarkovsky effect. Furthermore, the limited detection of the Yarkovsky effect when using only the *Gaia* data highlights the limitations of the current catalog and emphasizes the need for the inclusion of other astrometry and/or extension of the observational arcs.

Estimation of the close-approach distances of NEOs demonstrates the importance of including the Yarkovsky effect for close-approach analysis. Several detections presented have a very high SNR (>10). However the Yarkovsky effect should be included in the long-term orbital evolution, close approach and Earth-impact analysis for all NEOs with reliable detections and SNR >3. The final *Gaia* catalog may provide a much higher number of high-SNR detections of the Yarkovsky effect. This would allow for the inclusion of the effect in routine close approach and impact analysis for higher number of NEOs.

*Acknowledgements.* The research leading to these results has received funding from the National Science Center, Poland, grant number 2022/45/N/ST9/01403 in the years 2023/2024. DO was supported by the National Science Center, Poland, grant number 2022/45/ST9/00267, PB was supported by Grant nr 2022/45/ST9/00267 and through the Spanish Government retraining plan Mara Zambrano 2021-2023 at the University of Alicante (ZAMBRANO22-04). This work has used data from the European Space Agency (ESA) *Gaia* mission https://www.cosmos.esa.int/gaia This work presents results from the European Space Agency (ESA) *Gaia* space mission. *Gaia* data are being processed by the *Gaia* Data Processing and Analysis Consortium (DPAC). The funding for the DPAC is provided by national institutions, in particular the institutions participating in the *Gaia* Multilateral Agreement (MLA). The *Gaia* mission website is https://www.cosmos.esa.int/gaia. The *Gaia* archive website is https://archives.esac.esa.int/gaia and was possible thanks to the open source OrbFit software http://adams.dm.unipi.it/orbfit/. This research has used data and/or services provided by the International Astronomical Union Minor Planet Center, The JPL Database On-Line Ephemeris System, and the *Gaia* Collaboration. The authors thank all observers who submitted data to the Minor Planet Center. During the preparation of this work, the authors used Writefull for Overleaf and chatGPT for language editing and to improve readability. After using this tool/service, the authors reviewed and edited the content as needed and assume full responsibility for the content of the publication. The authors declare no competing interests.

## Appendix A: Estimated Yarkovsky effect

Table A.1: Estimated Yarkovsky effect for all accepted cases, orbital, physical parameters and the number of observations used for the orbit determination.

| Num | Name | $A_2$ $10^{-15}$ au d$^{-2}$ | SNR | H mag | Tax. | D km | $N_{MPC}$ | $N_{FPR}$ | $N_{radar}$ |
|---|---|---|---|---|---|---|---|---|---|
| 138947 | 2001 BA40 | -40.20±10.13 | 3.97 | 18.60 | S | 0.44±0.10 | 531 | 39 | 0 |
| 311554 | 2006 BQ147 | -31.86±5.46 | 5.83 | 18.70 | S | 0.38±0.02 | 507 | 60 | 2 |
| 38086 | Beowulf | -30.40±6.08 | 5.00 | 17.29 |  | 0.69±0.16 | 917 | 171 | 0 |
| 163000 | 2001 SW169 | -29.84±4.51 | 6.62 | 19.20 | S | 0.45±0.18 | 1167 | 106 | 0 |
| 33342 | 1998 WT24 | -22.87±1.78 | 12.85 | 17.90 | K | 0.40±0.06 | 1889 | 99 | 18 |
| 192563 | 1998 WZ6 | -22.39±5.75 | 3.89 | 17.25 | V | 1.20±0.04 | 1087 | 172 | 0 |
| 345705 | 2006 VB14 | -18.43±3.99 | 4.62 | 18.50 | Q | 0.40±0.10 | 1324 | 117 | 2 |
| 85236 | 1993 KH | -17.75±3.89 | 4.56 | 18.71 | S | 0.61±0.02 | 767 | 124 | 2 |
| 52750 | 1998 KK17 | -16.38±5.20 | 3.15 | 16.45 | V | 1.05±0.02 | 1439 | 232 | 0 |
| 7335 | 1989 JA | -16.19±3.98 | 4.07 | 17.00 |  | 0.73±0.02 | 2044 | 80 | 10 |
| 24761 | Ahau | -14.72±4.79 | 3.07 | 17.30 | C | 1.17±0.04 | 819 | 147 | 0 |
| 162181 | 1999 LF6 | -14.43±3.30 | 4.37 | 18.20 | S | 0.73±0.02 | 1725 | 67 | 3 |
| 85953 | 1999 FK21 | -14.23±1.22 | 11.66 | 18.10 | S | 0.63±0.13 | 1299 | 66 | 0 |
| 2063 | Bacchus | -13.25±2.10 | 6.31 | 17.25 | Q | 1.03±0.03 | 893 | 58 | 13 |
| 188174 | 2002 JC | -12.10±2.87 | 4.22 | 17.33 |  | 1.16±0.04 | 378 | 84 | 4 |
| 2062 | Aten | -11.58±0.67 | 17.29 | 17.10 | Q | 0.77±0.03 | 1028 | 234 | 8 |
| 1865 | Cerberus | -11.02±2.15 | 5.13 | 16.73 | Q | 1.20±0.06 | 2123 | 230 | 0 |
| 10563 | Izhdubar | -10.90±2.54 | 4.29 | 16.72 | Q | 1.53±0.06 | 627 | 142 | 0 |
| 7822 | 1991 CS | -9.31±1.83 | 5.09 | 17.33 | S | 1.44±0.01 | 1941 | 167 | 5 |
| 161989 | Cacus | -9.09±1.28 | 7.10 | 17.20 | S | 0.68±0.03 | 1658 | 149 | 0 |
| 164121 | 2003 YT1 | -6.41±0.98 | 6.54 | 16.20 | V | 1.14±0.12 | 1313 | 279 | 9 |
| 3103 | Eger | -7.69±1.86 | 4.14 | 15.19 | S | 1.24±0.06 | 4495 | 170 | 5 |
| 7341 | 1991 VK | -7.11±1.27 | 5.60 | 16.70 | Q | 0.79±0.04 | 2297 | 69 | 14 |
| 68950 | 2002 QF15 | -6.95±0.77 | 9.03 | 16.36 | S | 1.13±0.04 | 2963 | 158 | 14 |
| 2100 | Ra-Shalom | -5.91±0.43 | 13.74 | 16.24 | K | 2.30±0.20 | 407 | 127 | 0 |
| 85989 | 1999 JD6 | -5.89±1.24 | 4.75 | 17.06 | K | 1.62±0.03 | 3035 | 142 | 16 |
| 3200 | Phaethon | -5.89±0.46 | 12.80 | 14.32 | B | 4.98±0.21 | 6531 | 480 | 9 |
| 29075 | 1950 DA | -5.75±0.84 | 6.85 | 17.10 | L | 1.35±0.14 | 993 | 58 | 13 |
| 1943 | Anteros | -5.26±1.29 | 4.08 | 15.60 | S | 2.31±0.05 | 4943 | 235 | 0 |
| 3753 | Cruithne | -4.94±1.46 | 3.38 | 15.53 | S | 1.83±0.08 | 1152 | 140 | 0 |
| 1685 | Toro | -4.87±0.41 | 11.88 | 14.33 | S | 3.60±0.20 | 3788 | 259 | 10 |
| 137924 | 2000 BD19 | -3.86±0.98 | 3.94 | 17.20 | V | 0.97±0.04 | 838 | 126 | 8 |
| 1862 | Apollo | -3.73±0.20 | 18.65 | 16.08 | Q | 1.55±0.07 | 2375 | 243 | 18 |
| 1566 | Icarus | -3.64±0.24 | 15.17 | 16.34 | S | 1.07±0.06 | 1404 | 103 | 24 |
| 3908 | Nyx | 7.10±1.36 | 5.22 | 17.33 | V | 1.00±0.15 | 2071 | 75 | 17 |
| 385186 | 1994 AW1 | 10.93±1.61 | 6.79 | 17.60 | L | 0.82±0.03 | 2557 | 113 | 4 |
| 162142 | 1998 VR | 12.18±3.91 | 3.12 | 18.70 | S | 0.55±0.19 | 626 | 7 | 0 |
| 6239 | Minos | 14.01±2.78 | 5.04 | 18.49 | S | 0.47±0.12 | 1200 | 87 | 4 |
| 99907 | 1989 VA | 17.02±1.00 | 17.02 | 17.90 | Q | 0.64±0.17 | 1050 | 70 | 0 |
| 141531 | 2002 GB | 26.12±2.90 | 9.00 | 19.15 |  | 0.30±0.01 | 754 | 205 | 0 |
| 87024 | 2000 JS66 | 27.55±3.81 | 7.23 | 18.70 | S | 0.31±0.06 | 2026 | 54 | 2 |
| 138852 | 2000 WN10 | 39.19±4.26 | 9.20 | 20.20 | S | 0.31±0.01 | 1172 | 16 | 0 |
| 10302 | 1989 ML | 55.20±3.27 | 16.88 | 19.40 | E | 0.24±0.04 | 1110 | 79 | 0 |

Table A.2: Estimated Yarkovsky effect based on the old observations. Where $A_2^*$ are the values without using the precovery or old separated observations.

| Num | Name | $A_2$ $10^{-15}$ au d$^{-2}$ | $A_2^*$ $10^{-15}$ au d$^{-2}$ | SNR | H mag | Tax. | D km | $N_{MPC}$±$N_{FPR}$ | $N_{radar}$ |
|---|---|---|---|---|---|---|---|---|---|
| 55408 | 2001 TC2 | -29.67±7.70 | -68.24±20.39 | 3.85 | 18.80 |  | 0.46±0.01 | 286±54 | 0 |
| 99935 | 2002 AV4 | -14.05±3.90 | -5.31±15.02 | 3.60 | 16.04 | L | 1.53±0.27 | 1772±151 | 0 |
| 172034 | 2001 WR1 | -10.98±3.04 | -54.62±9.09 | 3.61 | 17.80 | S | 0.63±0.02 | 831±58 | 0 |
| 68350 | 2001 MK3 | -8.63±2.77 | -16.22±18.36 | 3.12 | 15.99 | S | 1.79±0.33 | 2074±383 | 0 |
| 363505 | 2003 UC20 | -7.78±0.56 | -19.17±6.27 | 22.93 | 18.40 | P | 1.88±0.01 | 594±175 | 6 |
| 11054 | 1991 FA | -4.83±1.03 | -48.66±19.0 | 16.90 |  | S | 1.41 | 0.05±1293 | 115±0 |
| 3554 | Amun | 6.20±1.87 | -14.79±6.70 | 3.32 | 15.82 | M | 3.19±0.03 | 1727±320 | 0 |

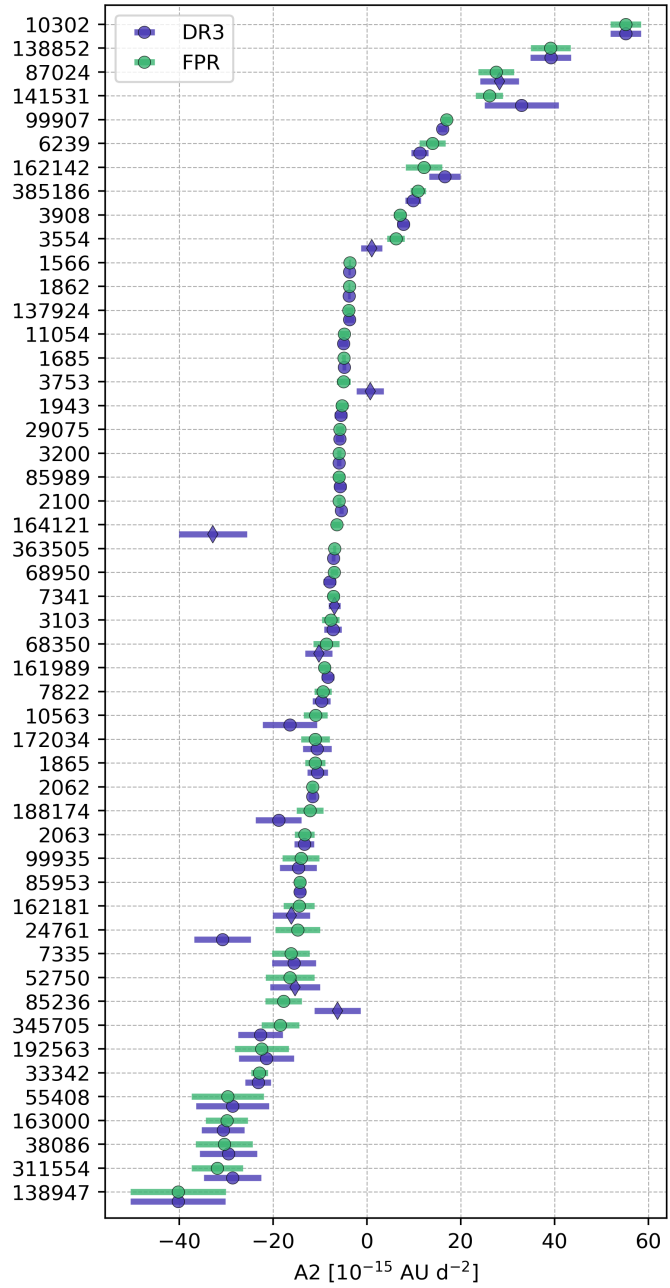

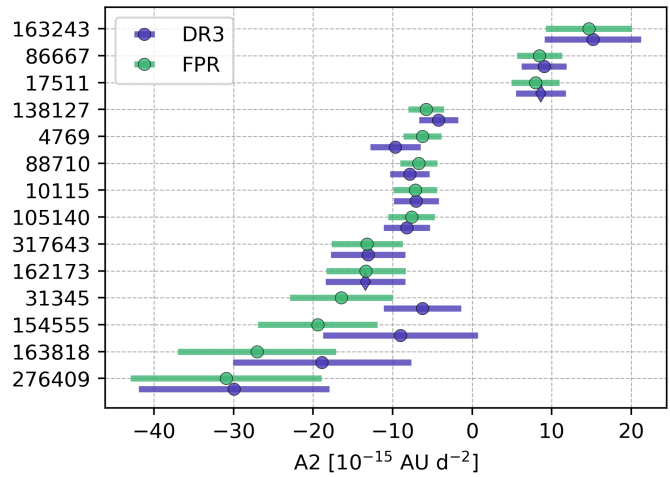

Fig. A.1: Comparison of the $A_2$ with 1-$\sigma$ uncertainty values computed using *Gaia* FPR and DR3. The top panel - accepted objects and the bottom panel - marginal objects. Diamond shaped purple points were rejected in *Gaia* DR3.